\magnification=\magstep1
\centerline{Compton Electrons and Electromagnetic Pulse in Supernovae and
Gamma-Ray Bursts}
\medskip
\centerline{J. I. Katz}
\centerline{Department of Physics and McDonnell Center for the Space
Sciences}
\centerline{Washington U., St. Louis, Mo. 63130}
\bigskip
\centerline{Abstract}
\medskip
{\narrower When gamma-rays emerge from a central source they may undergo Compton
scattering in surrounding matter.  The resulting Compton-scattered electrons
radiate.  Coherent radiation by such Compton electrons follows nuclear
explosions above the Earth's atmosphere.  Particle acceleration in
instabilities produced by Compton electron
currents may explain the radio emission of SN1998bw.  Bounds on coherent
radiation are suggested for supernovae and gamma-ray bursts; these bounds
are very high, but it is unknown if coherent radiation occurs in these
objects.\par}
\bigskip
High altitude (exoatmospheric) nuclear explosions are well known to produce
striking electromagnetic phenomena on the surface of the Earth.  These
phenomena, termed HEMP (High altitude ElectroMagnetic Pulse) or EMP (Karzas
and Latter 1962, 1965) occur when prompt gamma-rays following nuclear
fission, radiative neutron capture or inelastic neutron scattering suffer
Compton scattering in the upper atmosphere.  The Compton electrons, with
energies typically $\sim 1$ MeV, are preferentially directed along the
direction of the incident gamma-rays, radially away from the gamma-ray
source, and move at a speed close to the speed of light.  They are deflected
by the geomagnetic field and radiate synchrotron radiation.  Because the
gamma-rays and Compton electrons are produced over an interval $< 10^{-7}$
sec, shorter than the characteristic gyroperiod of the radiation ($\sim
10^{-6}$ sec, allowing for the relativistic energy of the electrons) this
radiation is coherent; it may be regarded as the effect of a continuously
distributed time-dependent current density, rather than as the radiation of
individual electrons.  The number of radiating electrons is very large so
that the currents and coherent radiation intensity are high, and are limited
by the condition that the radiation field not exceed the geomagnetic field,
for the radiation field of the Compton current acts to screen the
geomagnetic field.  In fact, the radiation may be crudely approximated as
the diamagnetic field exclusion by the conducting swarm of Compton
electrons.  In atmospheric EMP the Compton electrons produce large numbers
of low energy electrons by collisional ionization, and the effects of these
electrons on the emergent radiation are the chief subject of the published
calculations.

Analogous phenomena may be produced by astronomical events.  Karzas and
Latter (1965) indicate a threshold gamma-ray fluence for the observation of
EMP of $\sim 10^{-6}$ erg/cm$^2$.  This is less than the fluence of many
observed gamma-ray bursts (GRB), in some cases by more than two orders of
magnitude.  However, GRB do not produce observable EMP in the Earth's
atmosphere because their emission occurs over a duration from milliseconds
to minutes, several orders of magnitude (even for the shortest GRB) longer
than the electrons' gyroperiod in the geomagnetic field.  As a result the
EMP, although coherent in the sense that many electrons radiate in phase, is
much reduced in amplitude because it is an incoherent average over a very
large number of cycles of electron gyromotion; the source current varies
slowly, with only a very small Fourier component at synchrotron frequencies;
in contrast, EMP from nuclear explosions is produced by a current
distribution which is essentially a Dirac $\delta$-function in time.
Incoherent radiation by the individual electrons does have the usual
synchrotron spectrum, but because these fields add incoherently the total
power is small.

It may also be possible for radiation, analogous to EMP, by Compton
electrons to be produced within distant astronomical objects rather than in the
Earth's atmosphere.  This distant radiation may, in principle, be detected
at Earth.  Unfortunately, known astronomical sources of gamma-rays are not
nearly as impulsive as nuclear bombs, so the resulting radiation cannot be
predicted to be coherent.  However, in some cases plasma phenomena may
unexpectedly produce intense coherent radiation, so this possibility cannot
be disregarded with confidence.  A familiar example is the radio emission of
pulsars, which is intense enough to observe only because, in a manner not
yet well understood, coherent radiation by clumps of electrons is produced.

Many supernovae produce large quantities of the radioactive isotope
$^{56}$Ni, which decays by electron capture with a 6.1 day half life to
$^{56}$Co, which in turn decays, 80\% by electron capture and 20\% by e$^+$
emission, with a 77 day half life to stable $^{56}$Fe.  These decays produce
gamma-rays of several energies; the most important and abundant are the
0.812 MeV gamma-ray produced in 85\% of $^{56}$Ni decays and the 0.847 MeV
and 1.24 MeV gamma-rays produced in 100\% and 66\%, respectively, of
$^{56}$Co decays.  These gamma-rays then undergo Compton scattering as they
travel through the expanding supernova debris.  In the very energetic
SN1998bw nearly 1 M$_\odot$ of $^{56}$Ni was prodced (Iwamoto, {\it et al.}
1998).  SN1998bw was remarkable for its unprecedentedly intense, and
double-peaked, radio emission, with evidence of a self-absorption frequency
of several GHz, declining with time, and large magnetic fields (Kulkarni,
{\it et al.} 1998).  

The characteristic time scales of the Compton currents are set by the
radioactive half lives and the hydrodynamic expansion time scale, each of
which is many days.  It is thus unlikely that there will be coherent
synchrotron emission, as in nuclear EMP.  However, other kinds of emission
are possible.  Katz (1999) suggested that as the Compton electrons propagate
into surrounding matter there will be counterstreaming plasma instabilities
which may accelerate a few electrons to the Lorentz factors $\sim 10^2$
required to explain the observed synchrotron radiation.  The
synchrotron photosphere advances at the speed of the mildly relativistic
Compton electrons, in agreement with the observed (Kulkarni, {\it et al.}
1998) Lorentz factor of expansion of the synchrotron source of 1.6--2.  This
model is consistent with the observed 
double-peaked time history of the radio intensity, for the first peak may be
associated with Compton electrons following $^{56}$Ni decay.  The second
peak may occur when a second front of more energetic Compton electrons,
produced by the
more energetic gamma-rays of $^{56}$Co decay, overtake the slower initial
wave of Compton electrons.  

The inferred magnetic fields in SN1998bw are $\sim 10^{-1}$ gauss.  Such
fields are remarkably large for such a large astronomical object, apparently
implying a large magnetic flux (although only the magnitude, and not the
sign or direction, of the magnetic field is inferred form the observations).
Despite this, the magnetic field may
contain only a small fraction of the supernova energy, although
this last conclusion is very sensitive to the expansion Lorentz factor of
the radio source.  The synchrotron radiation by the Compton electrons is
therefore expected to be at unobservably low frequencies $\sim 1$ MHz.  The
observed radio emission implies a power law electron distribution function
extending to much higher energies.  There is no evidence of coherent
emission, nor would it be expected from electrons accelerated by a local plasma
instability driven by a smoothly varying flux of Compton electrons.  

An upper bound on the power radiated in coherent EMP by a source of size
$R$ is given by the condition that the radiation field not exceed the
background magnetic field $B$:
$$P < {B^2 \over 8 \pi} 4 \pi R^2 c \sim 10^{41} \left({B \over 0.1\,
{\rm gauss}}\right)^2 \left({R \over 3 \times 10^{16}\,{\rm cm}} \right)^2
{\rm erg/sec}.$$
This is about two orders of magnitude in excess of the observed (presumably
incoherent) radio luminosity of SN1998bw.  It is possible for the incoherent
synchrotron radiation power to exceed this bound on the coherent power; this
would correspond to the condition that synchro-Compton radiation exceed 
synchrotron radiation in power, which violates no fundamental condition,
although it is not often observed.

The plasma physics of GRB is more complex.  Their multi-peaked temporal
structure implies the existence of many interacting relativistic shells.  In
order to explain the dissipative interaction of shells with each other (and
with an external medium) there must be collective plasma interaction.  In
order to accelerate electrons to energies sufficient to produce the
observed radiation by the synchrotron process there must be at least
approximate electron-ion equipartition, again by some ill-understood
collective process.  Coherent emission can only be a speculation, but it is
worthwhile to bound it.

In order to bound the possible intensity of coherent EMP from GRB it is
necessary to bound their magnetic fields and dimensions.  The dimensions can
be reasonably, but very roughly, estimated from the observed time scales of
variations and Lorentz factors inferred by a variety of complex arguments.
The magnetic fields are, however, nearly unknown.  An earlier estimate (Katz
1994) of coherent EMP assumed microgauss interstellar fields, and concluded
that even if coherent EMP were produced it would be unobservably weak.  That
estimate might be applicable if GRB involved only an ``external'' shock
between the relativistic debris and the interstellar medium, but even in
that case the appropriate field may be a turbulently amplified field in the
shocked interstellar medium, possibly orders of magnitude higher.  It is now
generally accepted that GRB involve ``internal'' shocks between debris flows
of differing Lorentz factors.  Then the appropriate magnetic fields may be
those of the shocked debris clouds, which may approach equipartition with
the particle kinetic energy, although this last value is controversial and
estimates differing by several orders of magnitude exist.  Alternatively,
it may be that GRB are powered by low frequency electromagnetic radiation
from a magnetized accretion disc (Katz 1997).  If this radiation is not
completely converted to pair gas the magnetic energy density will be a
significant fraction of the relativistic hydrodynamic energy density in the
wind.  In either case, the upper bound on the coherent EMP may approach the
GRB power itself, although it is impossible to predict how the close the
actual coherent power (if there is any at all) comes to this bound.
Characteristic frequencies are also very uncertain, but typically of order
GHz.

An additional, and stricter, bound may be obtained by noting that the
radiating matter in GRB models is optically very thin, with Compton optical
depths typically $\sim 10^{-6}$, although dependent on very uncertain
parameters.  This sets an upper bound on the efficiency of conversion of
photon energy in Compton scattering.  This upper bound is further reduced
because in most models the photons have energy $\ll m_e c^2$ in the
co-moving frame, so only a small fraction of their energy is transferred to
the electron in a scattering event.  This suggests, very crudely, an overall
efficiency bound of $< 10^{-9}$.

The conditions for coherent EMP may be met in the evaporation of small
primordial black holes, for their final burst of gamma-rays is very brief.
The difficulty (aside from the very speculative assumption that such objects
exist) is in the astronomical environment.  In intergalactic or other nearly
empty space the Compton scattering length is very long (a problem exacerbated
by the decline of the Klein-Nishina cross-section with energy) and the
production of Compton electrons is negligible.  If the black hole is
captured by dense matter, and is found within a star, planet, or similar
object, then no radiation is observable because of absorption by the matter.
The most favorable conditions may be those of a black hole in a giant
molecular cloud, protostar, or a similar object, with column densities of
order the Compton scattering length, but the expected event rates are very
low, if not zero.  For such favorably located small black holes, detection of
their EMP may be more sensitive than direct observation of their gamma-rays.
\bigskip
\centerline{References}
\medskip
\parindent=0pt
Iwamoto, K., {\it et al.} 1998 {\it Nature} {\bf 395}, 672.

Karzas, W. J. and Latter, R. 1962 {\it J. Geophys. Res.} {\bf 67}, 4635

Karzas, W. J. and Latter, R. 1965 {\it Phys. Rev.} {\bf B137}, 1369.

Katz, J. I. 1994 {\it Ap. J.} {\bf 422}, 248.

Katz, J. I. 1997 {\it Ap. J.} {\bf 490}, 633.

Katz, J. I. 1999 {\it Ap. J.} submitted (astro-ph/9904053).

Kulkarni, S. R., {\it et al.} 1998 {\it Nature} {\bf 395}, 663.
\end